\documentclass[conference,letterpaper]{IEEEtran}

\usepackage[utf8]{inputenc}
\usepackage[T1]{fontenc}
\usepackage{url}                      
\usepackage{hyperref}
\usepackage{bm}
\usepackage{ifthen}
\usepackage{cite}

\usepackage{amssymb}
\usepackage[cmex10]{amsmath} 

\usepackage{graphicx}
\usepackage{color}

\usepackage{subfigure}
\usepackage{bbm}

\begin{document}
\title{Asymptotic Analysis on LDPC-BICM Scheme for Compute-and-Forward Relaying} 

\author{%
   \IEEEauthorblockN{Satoshi Takabe\IEEEauthorrefmark{1},
                        Tadashi Wadayama\IEEEauthorrefmark{1}, and 
                        Masahito Hayashi\IEEEauthorrefmark{2}\IEEEauthorrefmark{3}\IEEEauthorrefmark{4}}
  \IEEEauthorblockA{\IEEEauthorrefmark{1}
  	Department of Computer Science, Nagoya Institute of Technology, \{s\_takabe, wadayama\}@nitech.ac.jp}
  \IEEEauthorblockA{\IEEEauthorrefmark{2}
  	 Graduate School of Mathematics, Nagoya University, masahito@math.nagoya-u.ac.jp}
  \IEEEauthorblockA{\IEEEauthorrefmark{3}
  	 Shenzhen Institute for Quantum Science and Engineering, Southern University of Science and Technology}
  \IEEEauthorblockA{\IEEEauthorrefmark{4}
  	 Centre for Quantum Technologies, National University of Singapore}                    
}

\maketitle

\begin{abstract}
The compute-and-forward (CAF) scheme has attracted great interests due to its high band-width efficiency on two-way relay channels.
In the CAF scheme, a relay attempts to decode a linear combination of transmitted messages from other terminals or relays.
It is a crucial issue to study practical error-correcting codes in order to realize the CAF scheme
 with low computational complexity.      
In this paper, we present an efficient bit-interleaved coded modulation (BICM) scheme  
 for the CAF scheme with phase shift keying (PSK) modulations.
In particular, we examine the asymptotic decoding performance of the BICM scheme with low-density parity-check (LDPC) codes
by using the density evolution (DE) method.
Based on the asymmetric nature of the channel model, we utilize the population dynamics method for the DE equations without the all-zero codeword assumption.
The results show that, for two-way relay channels with QPSK and 8PSK modulations, 
the LDPC-BICM scheme {provides higher achievable rate}
compared with an alternative separation decoding scheme.
\end{abstract}

\section{Introduction}

Reliability and band-width efficiency of relays have been crucial issues in {relay-based} wireless communications 
 including satellite communications and mobile communications.
Recently, the compute-and-forward (CAF) scheme~\cite{Nazer11} attracts great interests because of its high band-width efficiency.
In the CAF scheme, a relay tries to decode a linear combination of the transmitted signals from other relays and terminals.
In the next time slot, the decoded message is transmitted 
to other relays and terminals.
It helps the wireless communication system to increase its band-width efficiency.
This concept is also crucial in the \emph{physical layer network coding}, which has been studied extensively~\cite{Katti08,Zhang09}.
In addition, the CAF scheme is effective for secure wireless communications~\cite{hwv}.

\begin{figure}[!t]
\centering
\includegraphics[width=0.375\linewidth]{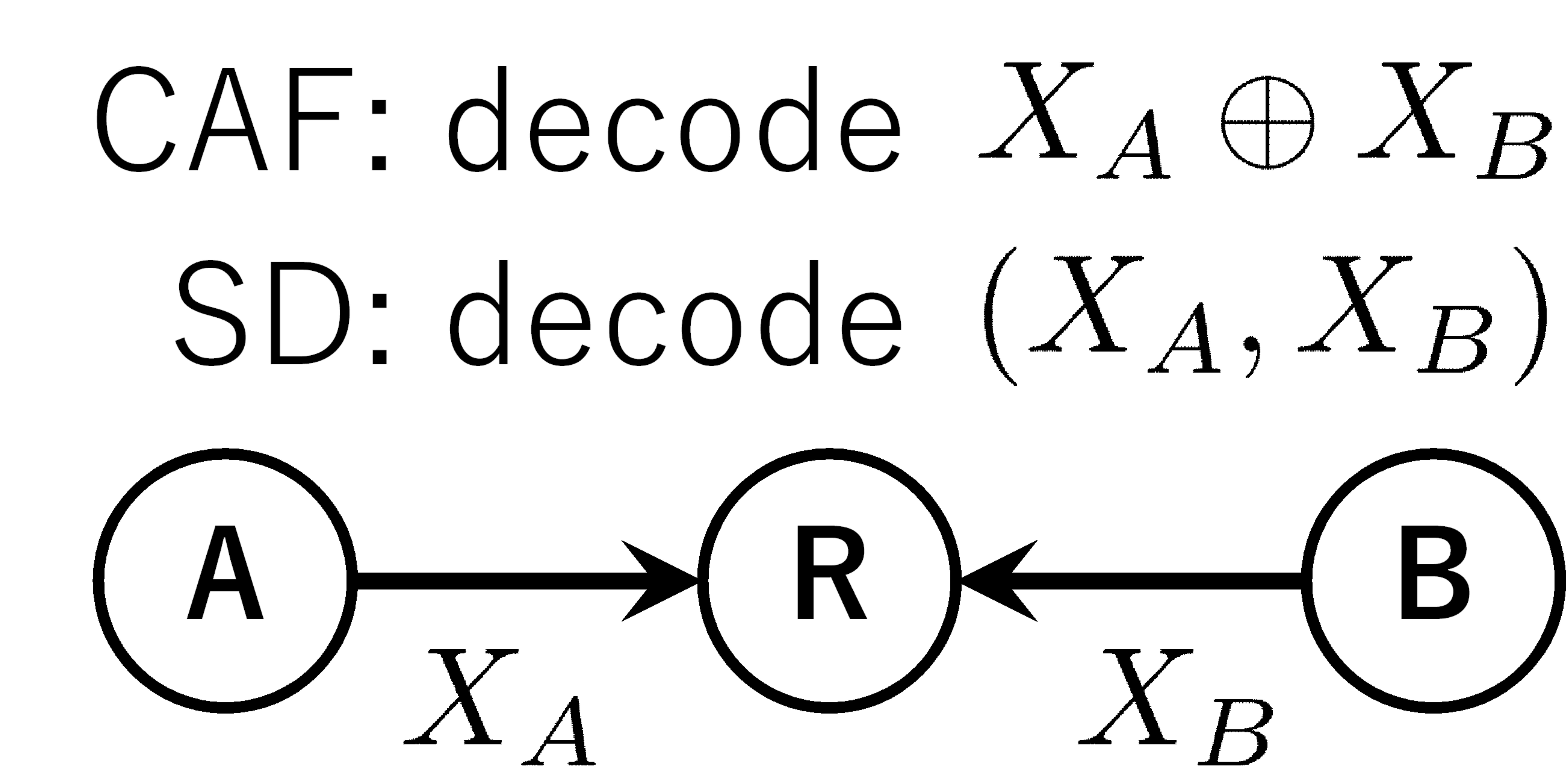}
\caption{Schematic diagram of the CAF and SD schemes in the MAC phase.
}\label{zu_i0}
\end{figure}

In order to investigate the performance limit of the CAF scheme, 
we discuss the performance under the framework of the two-way relay channel.
In the two-way relay channel {(Fig.~\ref{zu_i0})}, 
two terminals named A and B try to exchange their messages $X_A$ and $X_B$, respectively,
through bi-directional connections to the relay R, instead of any direct connection to each other.
It has two phases.
First, the two terminals simultaneously send their messages to R, which is called 
the multiple-access (MAC) phase.
Second, the relay R sends the received information to 
both terminals A and B.
In the second phase, it is sufficient that the relay R 
broadcasts the modulo sum $X_A\oplus X_B$ to both terminals A and B because
they have their own original messages.
Therefore, the relay R may decode only the modulo sum $X_A\oplus X_B$ in the MAC phase, which is called the CAF scheme.
In contrast, Yedla et al. \cite{Yedla09} proposed an efficient decoding strategy for decoding 
both messages $X_A$ and $ X_B$ in the MAC phase, which is called the MAC separation decoding (SD) scheme 
(see Fig.~\ref{zu_i0}). 
The SD scheme has an advantage in directly decoding the messages without loosing information as in the CAF scheme
while a decoder for the CAF scheme is much simpler than that for the SD scheme.
It is thus crucial to analyze their decoding performance in the MAC phase
 to realize an efficient and practical relaying technique with high reliability and low {complexity}.
 
{The key to realize a practical CAF scheme is a proper choice of error-correcting codes.
The low-density parity-check (LDPC) code} is a leading candidate due to its high reliability against
 various channel models including an additive white Gaussian noise (AWGN) channel~\cite{MacKay99}. 
Sula et al. proposed the LDPC coding technique for the CAF scheme with the binary phase shift keying (BPSK) modulation and evaluated its 
decoding performance by numerical simulations~\cite{Sula17}.
The authors {theoretically} analyzed the LDPC codes and its spatial coupling coding 
by the \emph{density evolution} (DE) method
 and found that they {exhibit higher reliability} than the SD scheme~\cite{tiwh,twh2}. 

{
For realizing a CAF scheme with non-binary modulations, lattice-based CAF methods based on PAM or QAM modulation
have been extensively studied~\cite{Nazer11,Feng}. 
This is because the additive group property of lattices is well matched to the CAF scheme.
Although the lattice-based CAF methods provide near optimal performance when SNR is high enough,
there remain two issues regarding its practical implementation. 
The first one is the decoding complexity.
In general, lattice decoding is computationally expensive and harder to implement compared with binary channel coding.
The second problem is that the lattice-based method cannot apply to a PSK or APSK modulation format,
which is often employed in satellite communications.
In order to find a possible solution for these problems,
we here discuss a bit-interleaved coded modulation (BICM) with $2^K$-PSK modulations
as the basis of a CAF scheme. 
In the BICM scheme, a bit interlearver is set between an encoder and a mapper to separate
 the coding and modulation~\cite{BICM}.
It remarkably reduces the computational cost of a decoder in spite of a small loss in decoding performance. 
Although the LDPC-BICM scheme for the CAF scheme is studied in~\cite{Du},
the analysis is only based on the numerical simulations. 
This means that theoretical analysis unveiling the achievable rate with belief propagation (BP) decoding remains open.
}


In this paper, we will present the asymptotic performance analysis of the LDPC-BICM sheme 
{in the MAC phase of the CAF scheme on the two-way relay channel.
First, we will propose a DE analysis of the LDPC-BICM scheme 
without any conventional approximations. 
Then, the achievable rate of of the LDPC-BICM scheme for the CAF scheme
 is obtained by the DE analysis, which offers theoretical comparison with the alternative SD scheme.
}

\section{LDPC-BICM Scheme and Density Evolution}\label{sec_normal}

{Before describing the main results, 
we first propose a novel DE analysis of the LDPC-BICM scheme. 
Although the LDPC-BICM scheme has been extensively studied by a DE analysis~\cite{BICM,Lei},
the analysis is usually based on some \emph{approximations}.
A well-known approximation is the \emph{all-zero codeword assumption}~\cite{Hou} in which 
a transmitter send an all-zero codeword as an instance of random codewords.
{Another approximation is the assumption on the distribution of extrinsic output values
in the \emph{extrinsic information transfer (EXIT) chart} analysis~\cite{Ash}.}
Unfortunately, these assumptions do not hold in general cases including the LDPC-BICM scheme
 on higher order modulations.
Here, we will propose an alternative DE analysis based on that for asymmetric channels~\cite{Wang05}.
}

For simplicity, we consider the LDPC-BICM scheme for a single-access channel model such as a complex AWGN (CAWGN) channel.
Let us consider a {binary} LDPC code $C\subset \mathbb{F}_2^{Kn}$ ($\mathbb{F}_2\triangleq \{0,1\}$).
At a transmitter, a message is encoded to ${x} = (x_s^i)^{i=1,\dots , n}_{s=1,\dots, K}\in C$.
In the BICM scheme, the transmitter uses a bit interleaver $\pi$ to remove correlations in the interleaved signal $\pi({x})$.
The signal $\pi({x})$ is then modulated by a mapper and transmitted through a channel.
A receiver attempts to demap and decode $\pi({x})$ from the received signal $y$, and 
detects the transmitted signal $x$ using the interleaver $\pi^{-1}$.
Assuming that the code length $n$ is sufficiently large and an interleaver $\pi$ is randomly generated,
each element of $\pi({x})$ is sufficiently uncorrelated and the receiver can decode $x_s^i$ \emph{one by one}.
It makes the structure of the decoder considerably simple because a standard {BP decoder for binary LDPC codes}
 is available only by replacing its log-likelihood ratio (LLR) calculation unit.
In fact, if  $p_{Y|\tilde{X}}(y|\tilde{x})$ is the conditional PDF
corresponding to the constellation map and channel model {for each $K$ bits $\tilde{x}\in \mathbb{F}_2^K$,
 the LLR for the $s$th bit in $\tilde{x}$ is given by}
\begin{align}
\tilde\lambda_s({y})&\triangleq \ln \frac{\tilde{L}_s[y|1]}{\tilde{L}_s[y|0]} \quad (s=1,\dots,K) 
, \label{eq_norm1}
\end{align}
where 
$\tilde{L}_s[y|u]$ ($u\in\mathbb{F}_2$) is the likelihood function of $\tilde{x}$ whose $s$th bit is $u$, which is defined by
\begin{equation}
\tilde{L}_s[{y}|u]\triangleq \frac{1}{2^{K-1}}\sum_{\tilde{x}:\tilde{x}_s=u}p_{Y|\tilde{X}}({y}|\tilde{x}). \label{eq_norm2}
\end{equation}

\begin{figure}[!t]
\centering
\includegraphics[width=0.86\linewidth]{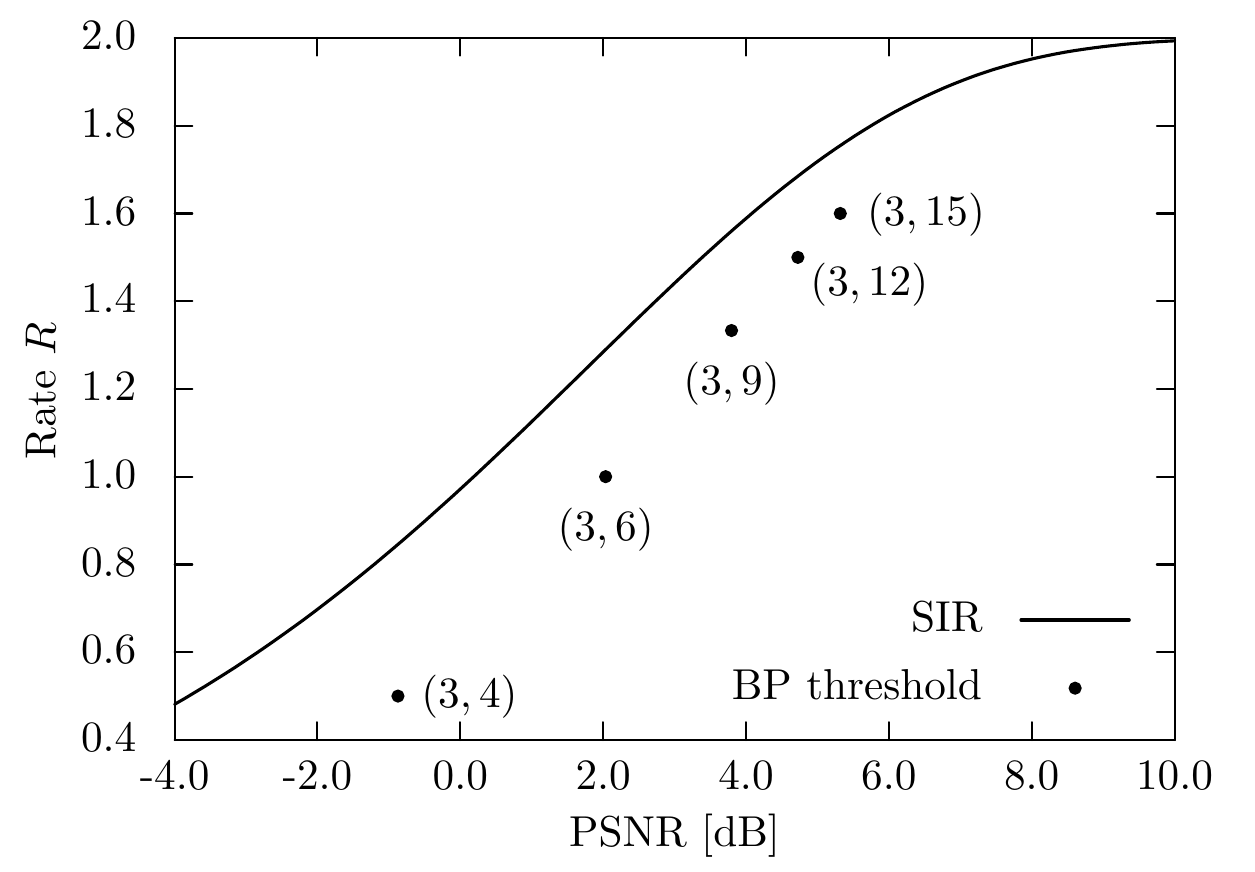}
\caption{BP thresholds of the LDPC-BICM scheme with the QPSK modulation as a function of the code rate $R$.
The solid line represents the SIR of the CAWGN channel.
}\label{zu_s1}
\end{figure}

The DE method is useful to analyze an asymptotic decoding threshold called BP threshold.
Now we introduce the DE method for the LDPC-BICM scheme {without conventional assumptions}.
Let us consider the $(d_v,d_c)$-regular LDPC-BICM scheme, where $d_v$ and $d_c$ represent 
the variable and check node degrees, respectively.
The conditional PDF $P^{(l)}(m|u)$ (resp. $Q^{(l)}(\hat{m}|u)$) denotes 
the PDF of a message $m$ from a variable node to a check node
(resp. $\hat{m}$ from a check node to a variable node) with a transmitted bit $u$ at the $l$th step.
Following the DE equations for binary asymmetric memoryless channels~\cite{Wang05}, we have
\begin{align}
P^{(l)}(m|u)&\!=\frac{1}{K}\sum_{s=1}^K
\int_{\mathbb{C}}\! d{y}\tilde{L}_s[{y}|u]\int\! \prod_{d=1}^{d_v-1}\! d\hat{m}^{d}Q^{(l-1)}\! (\hat{m}^{d}|u)\nonumber\\
&\times\delta\left(m-\tilde\lambda_s({y})-\sum_{d=1}^{d_v-1}\hat{m}^{d}\right), \label{eq_d3}\\
Q^{(l)}(\hat{m}|u)&\!=\!\frac{1}{2^{d_c-2}}\sum_{(u^{d})\in U_{d_c}^u}
\int \prod_{d=1}^{d_c-1}d{m}^{d}P^{(l)}({m}^{d}|u^{d})\nonumber\\
&\times\!\delta\left(\hat{m}\!-\!2\tanh^{-1}\left[\prod_{d=1}^{d_c-1}\tanh\left(\frac{{m}^{d}}{2}\right)\right]\right),\! \label{eq_d4}
\end{align}
where  
$
U_D^u\triangleq\left\{(u^{d})\in \mathbb{F}_2^{D}:\bigoplus_{d=1}^{D}u^{d}=0, u^{D}=u\right\}
$.
To derive these equations, we use a condition that $\pi$ is uniformly chosen from all the possible permutations.
{We assume that the bit index $s$ in~(\ref{eq_norm1}) becomes an independent random variable
 in the large-$n$ limit due to the random permutation. 
 This is a crucial assumption for the above DE analysis.}

\begin{figure*}[t]
  \centerline{\subfigure[QPSK]{\includegraphics[width=2.8in]{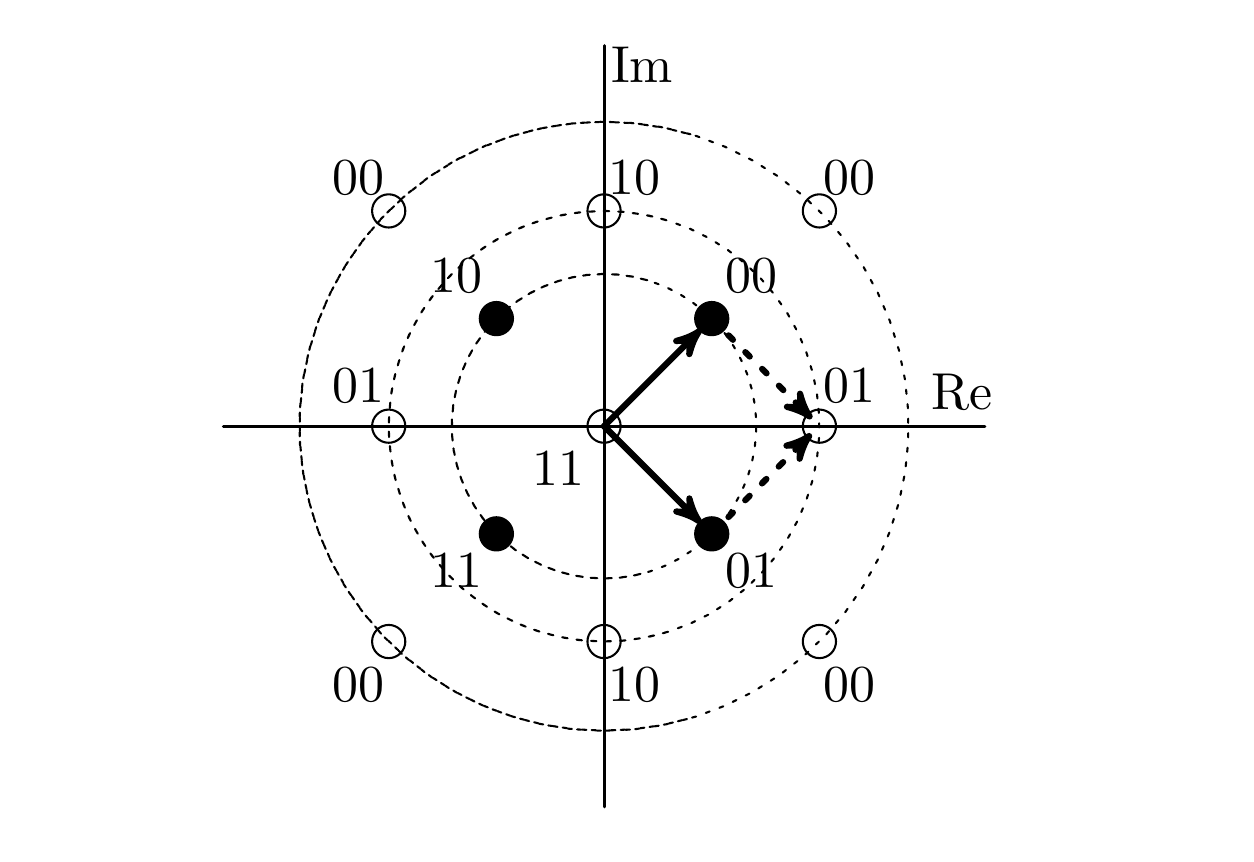}
       \label{zu_c1}}
     \hfil
     \subfigure[$8$PSK]{\includegraphics[width=2.8in]{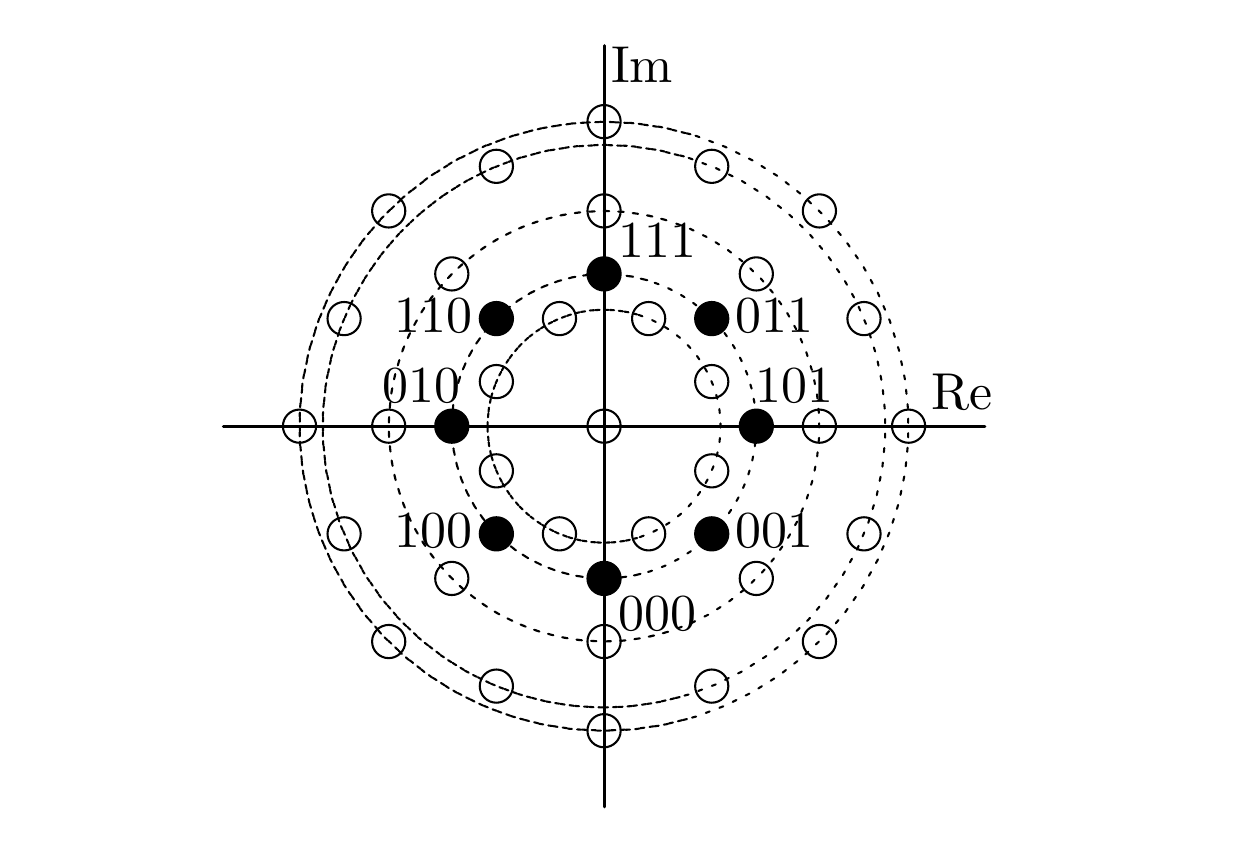}
       \label{zu_c2}}}
\caption{Constellation diagram of (a) QPSK modulation and (b) $8$PSK modulation.
Open and closed points are signal points at terminals and the relay, respectively.
Each label represents a bit sequence corresponding to a signal point.
In Fig.~\ref{zu_c1}, the arrows illustrate that the received signal point corresponding to label $01$ is generated by a pair of transmit signals whose labels are $00$ and $01$.
}
   \label{zu_c0}
\end{figure*}
We then employ the population dynamics (PD) method to solve the DE equations efficiently.
The method is popular in statistical physics~\cite{Mezard} and have been applied to the DE analysis~\cite{tiwh,twh2}.
Note that a conventional numerical analysis based on the fast Fourier transformation is also available
but the PD method is more straightforward for {evaluating~(\ref{eq_d3}), (\ref{eq_d4})}.
In the PD method, the PDFs $P(\cdot|u)$ and $Q(\cdot|u)$ ($u\in\mathbb{F}_2$) are approximated
by histograms of $N$ samples. 
The parameter $N$ is called the population size and the DE equations are exactly solved in the large-$N$ limit.
Each sample is recursively updated by an update rule written in a delta function $\delta(\cdot)$ 
in~(\ref{eq_d3}) or~(\ref{eq_d4}) up to the $T$th iteration step.
More detailed description is found in~\cite{tiwh}.
In the BICM scheme, it is necessary to add a sampling step of $s$ in~(\ref{eq_d3}) to the PD method.
The LLR function $\tilde L_s[y|u]$ is chosen by a sampled value of $s$.

We now demonstrate the DE analysis of the LDPC-BICM scheme on a CAWGN channel.
We use the QPSK modulation ($K\!=\!2$) with the Gray mapping.
Figure~\ref{zu_s1} shows the BP thresholds as a function of the code rate $R=K(1-d_v/d_c)$.
In the PD method, we set $N\!=\!10^5$ and $T\!=\!2000$, 
which is sufficiently large to evaluate a BP threshold for an AWGN channel
with the BPSK modulation. 
For comparison, we show the symmetric information rate (SIR) defined as the mutual information between 
the transmitted signal and received signal assuming that the transmitted word is chosen uniformly.
We find that the DE analysis reasonably obtains the BP thresholds of the LDPC-BICM scheme. 
It is emphasized that the result is asymptotically exact as shown in~\cite{Wang05} because the analysis
 directly treats the asymmetric channel. 
{This is the first case of evaluating BP thresholds by DE equations without any conventional approximations as far as the authors 
aware of.}

\section{{CAF and SD} Schemes on MAC Phase}
In this section, we define the CAF and SD schemes on the MAC phase of the two-way relay channel.
Let us define a $2^K$-PSK modulation 
by a constellation mapper $\mathcal{M}:\mathbb{F}_2^K\rightarrow \mathbb{C}$.
We assume that two terminals A and B use the QPSK and $8$PSK modulations
 in Fig.~\ref{zu_c0}.
For each time slot $t(=1,\dots,n)$, A and B respectively transmit their messages
$X_A^{(t)},X_B^{(t)}(\in\mathbb{F}_2^{K})$.
The two-way relay channel~\cite{Noori,Wu} is then defined by 
\begin{equation} \label{channel_model}
	Y^{(t)} = \mathcal{M}(X_A^{(t)}) + \mathcal{M}(X_B^{(t)})e^{i\theta} + W^{(t)},
\end{equation}
where $Y^{(t)}(\in \mathbb{C})$ is a received signal at the relay R at time slot $t$
and $\theta$ is the phase difference between two terminals.
Here, we assume that the perfect phase synchronization and perfect power control 
are available at R and  $\theta$ can be tuned to an arbitrary value.
Each element of $W^{(t)}$ is an i.i.d. complex Gaussian random variable with 
zero mean and variance $\sigma^2$. Its PDF is given by $F_c(w; 0, \sigma^2)$,
where
\begin{equation}
F_c(w; \mu, \sigma^2) \triangleq \frac{1}{{\pi \sigma^2}} 
\exp \left( -\frac{|w-\mu|^2}{\sigma^2} \right)\quad (w,\mu\in\mathbb{C})
\end{equation}
We use peak signal-to-noise ratio (PSNR) defined by 
{$\mathrm{PSNR}\triangleq 10\log_{10}(\max_{r\in \mathcal{M}(X): X\in\mathbb{F}_2^K}|r|^2/\sigma^2)$ [dB]}.

In the SD scheme,
both terminals transmit their message with their encoding
and the relay R decodes them simultaneously.
In the CAF scheme, 
the relay R decodes only their modulo sum.
In particular, in the CAF scheme with a degraded channel,
to detect the transmitted signals from A and B, 
the relay R adapts the decoding scheme with the degraded channel, which is defined as follows. 
When the input signal is $Z \in \mathbb{F}_2^K$, 
the output $Y$ of the degraded channel is defined by
\begin{equation} \label{deg-channel_model}
	Y \triangleq \mathcal{M}(X_A) + \mathcal{M}(Z \oplus X_A )e^{i\theta} + W,
\end{equation}
where
$X_A$ is the random variable {uniformly chosen from $\mathbb{F}_2^K$},
 and $W$ is the complex Gaussian random variable with 
zero mean and variance $\sigma^2$.
The PDF $p_{Y|Z}$ is given by
\begin{equation}
p_{Y|Z}(y|z) 
\triangleq 
\frac{1}{2^K}\sum_{\substack{x_A,x_B\in \mathbb{F}_2^K:\\ x_A\oplus x_B=z}}
F_c(y;\mathcal{M}(x_A)+ \mathcal{M}(x_B)e^{i\theta},\sigma^2)
\label{eq_sd_caf4b}.
\end{equation}
{In other words, for each $z\in \mathbb{F}_2^K $, we have 
the \emph{received constellation} at R~\cite{Dana,Noori}, which is defined by} 
$M_{2,\theta}^z\triangleq \{\mathcal{M}(x_A)+ \mathcal{M}(x_B)e^{i\theta}: x_A\oplus x_B=z\}$.
The total received constellation 
$M_{2,\theta}\triangleq\cup_{z\in\mathbb{F}_2^K}M_{2,\theta}^z$
is represented by open points in Fig.~\ref{zu_c0} .

\section{Asymptotic Analysis of LDPC-BICM Scheme for {CAF Scheme}}\label{sec_de}

Now we turn to the main results on the LDPC-BICM scheme for the CAF scheme with the degraded channel.
In the BICM scheme, we assume that {two terminals use the same {binary} LDPC code $C$ and 
both terminals and the relay use the same random interleaver $\pi$.}
Then, the LDPC-BICM scheme is defined as described in Section~\ref{sec_normal} because of the linearity of LDPC codes.
By using the PDF $p_{Y|Z}$ in \eqref{eq_sd_caf4b},
the LLR and the likelihood function 
read
\begin{align}
\lambda_s({y}) &\triangleq \ln \frac{L_s[y|1]}{L_s[y|0]}, \label{eq_bicm1}
\\
L_s[{y}|u]& \triangleq \frac{1}{2^{K-1}}\sum_{{z}:z_s=u}p_{Y|Z}({y}|{z}), \label{eq_bicm2}
\end{align}
The received signal is decoded as explained in Section~\ref{sec_normal}. 
The only difference from Section~\ref{sec_normal}
is to replace the LLR~(\ref{eq_norm1}) and the likelihood function (\ref{eq_norm2}) 
by (\ref{eq_bicm1}) and \eqref{eq_bicm2},
respectively.
Consequently, the LDPC-BICM scheme for the CAF scheme has an advantage in a simple decoding structure.
Although the BICM scheme is also applicable to the SD scheme, its BP decoder is rather complicated 
{because it decodes} a pair of transmitted codewords~\cite{Yedla}.
In addition,
the DE analysis in Section~\ref{sec_normal} is easily extended to the {present} LDPC-BICM scheme.
In fact,  we have the same DE equations with~~(\ref{eq_d3}) and (\ref{eq_d4}) 
with the above replacement.
We can apply the PD method to the DE equations to estimate BP thresholds.

\begin{figure}[!t]
\centering
\includegraphics[width=0.86\linewidth]{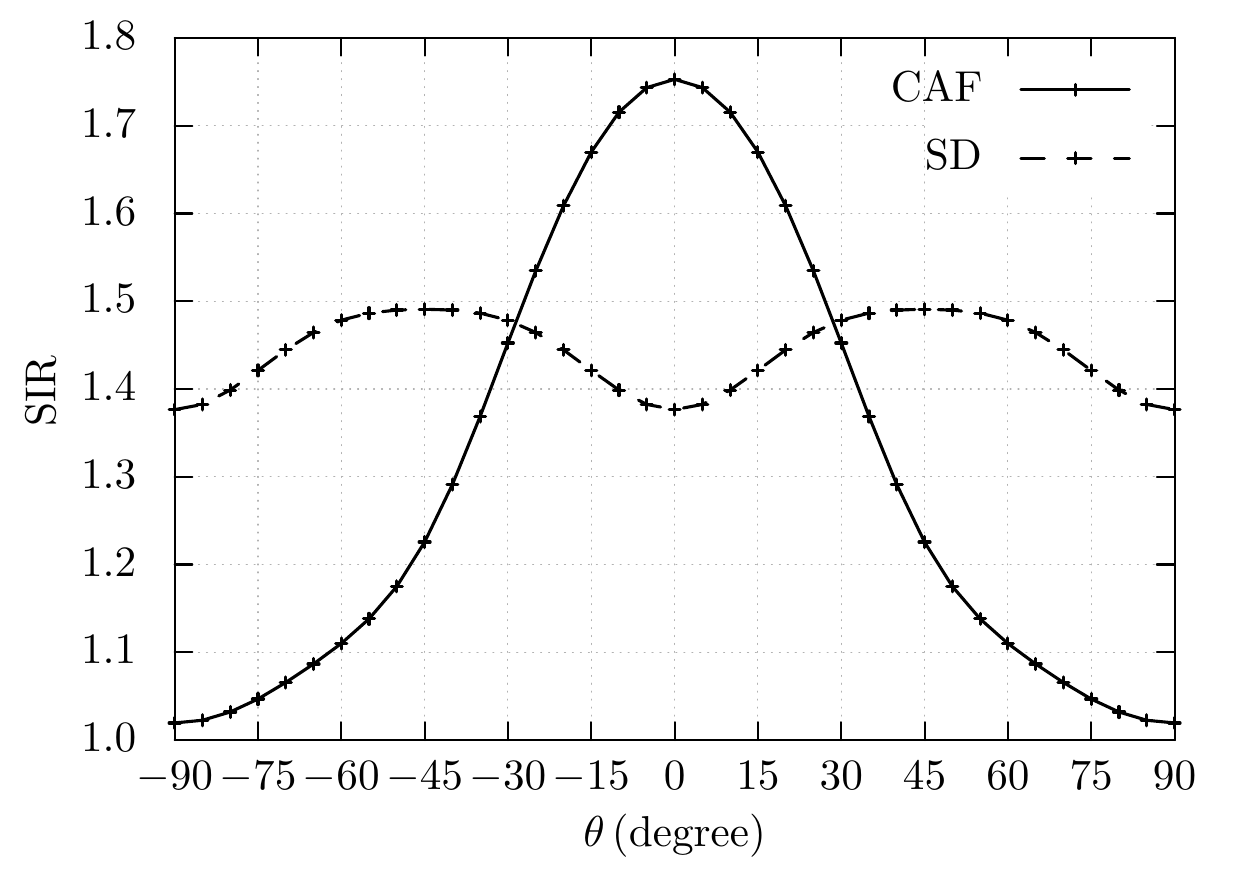}
\caption{The SIRs of the CAF and SD schemes for the QPSK modulation (PSNR$=6$[dB])  as a function of the phase difference $\theta$.
}\label{zu_s2}
\end{figure}

\begin{figure}[!t]
\centering
\includegraphics[width=0.86\linewidth]{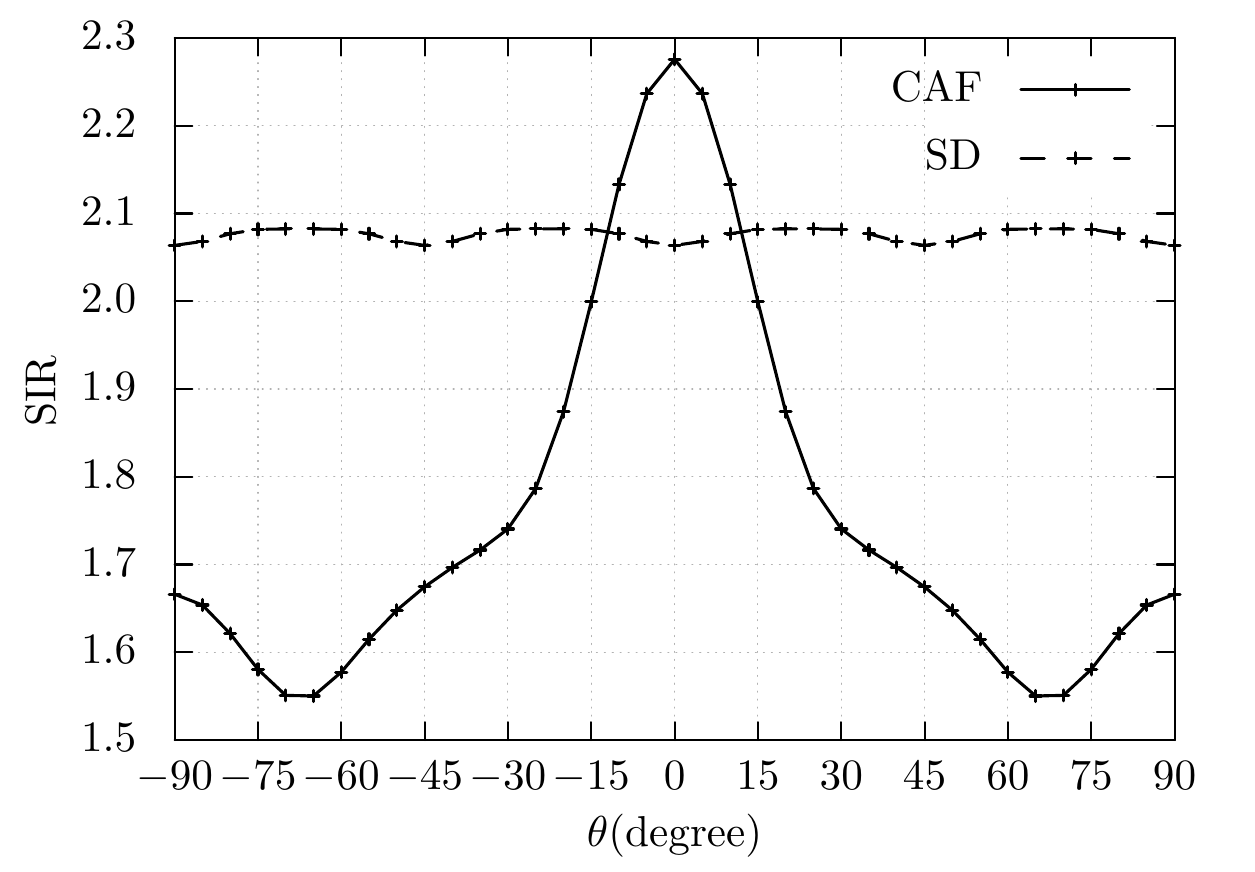}
\caption{The SIRs of the CAF and SD schemes for the 8PSK modulation (PSNR$=10$[dB])  as a function of the phase difference $\theta$.
}\label{zu_s3}
\end{figure}

To compare the CAF and SD schemes, we utilize their SIRs, 
which express the asymptotic transmission rates of random linear codes. 
By using the uniform distribution $P_Z$ on $\mathbb{F}_2^K$
and the PDF $p_{Y|Z}$ given in \eqref{eq_sd_caf4b},
the SIR of the CAF scheme with the degraded channel is given by
\begin{align}
I(Y;Z) 
=& \sum_{z\in \mathbb{F}_2^K}P_Z(z)\int_{\mathbb{C}} dy p_{Y|Z}(y|z)\log_2 p_{Y|Z}(y|z)
\nonumber\\
&  -\int_{\mathbb{C}} dy p_Y(y) \log_2 p_Y(y),
\label{eq_sir_caf}
\end{align}
while that of the SD scheme reads
\begin{align}
\frac{1}{2}I(Y;X_A, X_B) 
=& \frac{1}{2}\int_{\mathbb{C}} dw F_c(w; 0, \sigma^2)\log_2 F_c(w; 0, \sigma^2)
 \nonumber\\
&-\frac{1}{2}\int_{\mathbb{C}} dy p_Y(y) \log_2 p_Y(y),
\label{eq_sd_caf}
\end{align}
where 
\begin{align*}
p_Y(y) 
\triangleq& \frac{1}{2^{2K}}
\sum_{x_A,x_B\in \mathbb{F}_2^K} 
& F_c(y;\mathcal{M}(x_A)+ \mathcal{M}(x_B)e^{i\theta},\sigma^2). \label{eq_sd_caf4d}
\end{align*}
{Note that the SIR of the CAF scheme depends on a labeling of a constellation map while that of the SD scheme does not.
It implies that an optimal labeling for the CAF scheme exists as studied in~\cite{Dana}.
In this paper, we use the constellations in Fig.~\ref{zu_c0} which show reasonably high SIRs 
because no labels are overlapped at signal points in their received constellations.}
As shown in~\cite{Wu}, SIRs depend on the phase difference $\theta$.
Figures~\ref{zu_s2},~\ref{zu_s3} respectively show the SIRs of  both schemes for {the QPSK and $8$PSK modulation in Fig.~\ref{zu_c0}  as a function of $\theta$.}
We find that the SIRs of the CAF scheme take maximum values at $\theta\!=\!0$.
In contrast, the SIR of the SD scheme is maximized at a nontrivial value of $\theta$ 
because received constellation points of different transmitted signal pairs, e.g., $(X_A,X_B)=(10,01), (01,10)$, are overlapped when $\theta=0$.
It shows that the value of $\theta$ is set properly when we compare both schemes.
In the following experiments, we set $\theta=\pi/4$ for the QPSK modulation and $\theta=\pi/8$ for the $8$PSK modulation.   

\begin{figure}[!t]
\centering
\includegraphics[width=0.86\linewidth]{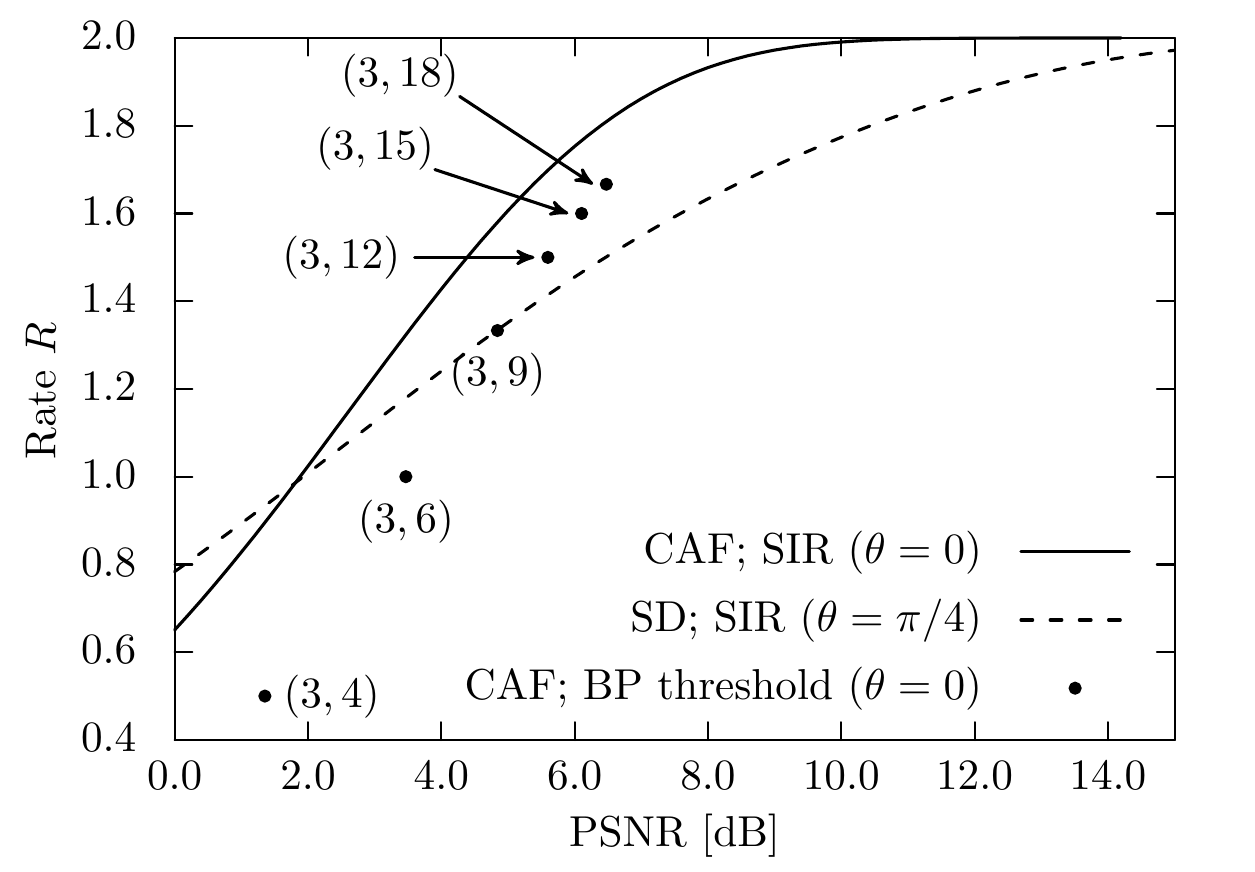}
\caption{The BP thresholds of the LDPC-BICM scheme (symbols) and SIRs of the CAF and SD schemes with the QPSK modulation (lines) as a function of the PSNR.
Each label represents $(d_v,d_c)$ of the regular LDPC code ensemble.
}\label{zu_s4}
\end{figure}

Figure~\ref{zu_s4} shows the BP thresholds of some LDPC-BICM schemes with the QPSK modulation.
We used $N\!=\!10^5$ and $T\!=\!2000$ in the PD method.
The figure also shows the SIRs of the CAF and SD schemes with appropriate $\theta$ to maximize them.
We confirm that, in terms of the SIR, the CAF scheme is superior to the SD scheme in the high rate regime as with the BPSK modulation~\cite{twh2}.
On the other hand, when the rate $R$ is below $1.0$, the SD scheme becomes effective because the CAF scheme is based on the degraded channel.
In addition, the BP thresholds of the LDPC-BICM shceme for the CAF scheme is superior to the SIR of the SD scheme.
For example, the BICM scheme with $(3,18)$-regular LDPC codes has about $2.0$ dB gain against the SIR of the SD scheme.

\begin{figure}[!t]
\centering
\includegraphics[width=0.86\linewidth]{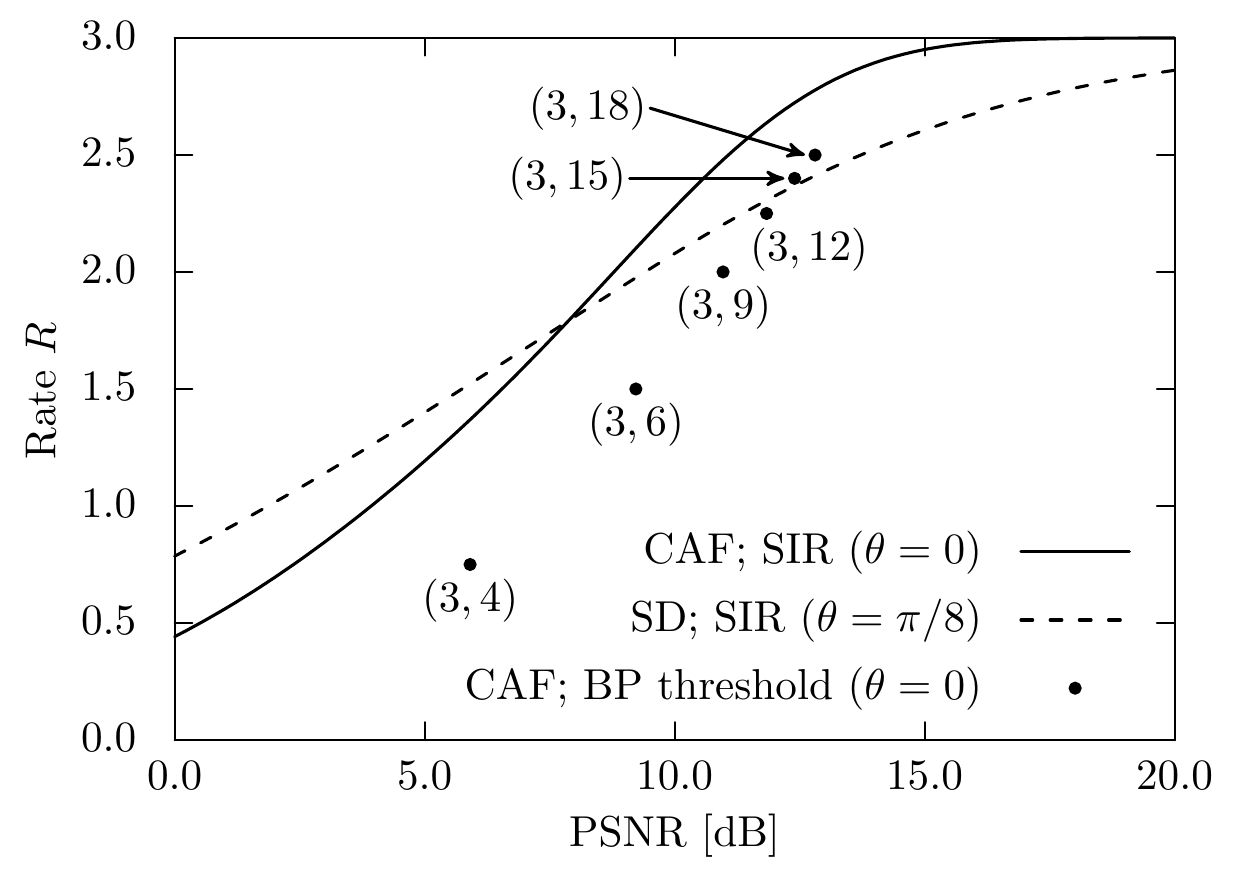}
\caption{The BP thresholds of the LDPC-BICM scheme (symbols) and SIRs of the CAF and SD schemes with the $8$PSK modulation (lines) as a function of the PSNR.
Each label represents $(d_v,d_c)$ of the regular LDPC code ensemble.
}\label{zu_s5}
\end{figure}

We also show the BP thresholds of LDPC-BICM schemes with the $8$PSK modulation in Fig.~\ref{zu_s5}.
The CAF scheme is superior to the SD scheme in terms of the SIR when the rate $R$ is larger than $1.8$.
For practical LDPC-BICM shcmes, the BICM scheme with $(3,18)$-regular LDPC codes has $1.0$ dB gain compared with the SIR of the SD scheme.  
It is emphasized that the gain becomes larger {if a practical error-correcting code is applied to the SD scheme.}

\section{Summary}

In this paper, we have studied the LDPC-BICM scheme for the CAF scheme with the degraded channel on the two-way relay channel.
We have investigated its asymptotic decoding performance by using a novel DE method.
The results show that the LDPC-BICM scheme exhibits higher reliability than the {alternative} SD scheme in the high rate regime.
It indicates that the LDPC-BICM scheme for the CAF scheme is  
an effective and practical coded modulation scheme to realize efficient and reliable relaying 
because of its simple decoding structure, low computational cost,  and {high achievable rate.}

\section*{Acknowledgment}

This work is supported
by JSPS Grant-in-Aid for Scientific Research (A) Grant Number 17H01280.



\begin{thebibliography}{99}


\bibitem{Nazer11} 
B. Nazer and M. Gastpar,
``Compute-and-forward: harnessing interference through 
structured codes,'' 
IEEE Trans. Inf. Theory, vol. 57, no. 10, pp. 6463-6486, Oct. 2011.


\bibitem{Katti08}
S. Katti,  H. Rahul,  W. Hu,  D.  Katabi,  M. Medard, and  J. Crowcroft,
``XORs in the air: practical wireless network coding, ''
IEEE/ACM Trans. Networking, vol. 16, no. 3, pp. 497-510, Jun. 2008.

\bibitem{Zhang09}
S. Zhang, and S.-C. Liew,
``Channel coding and decoding in a relay system operated with physical-layer network coding, ''
IEEE J. Select. Areas in Commun., vol. 27, no. 5, pp. 788-796, Jun. 2009.

\bibitem{hwv}
M. Hayashi, T. Wadayama, and A. Vazquez-Castro, ``Secure Computation-and-Forward with Linear Codes,''
\textit{Inf. Theory Workshop}, Guanzhou, China, 2018; arXiv:1804.10729.



\bibitem{Yedla09} A. Yedla, H. D. Pfister and K. R. Narayanan, 
``Can iterative decoding for erasure correlated sources be universal?'' 
\textit{2009 47th Annual Allerton Conf. Comm., Control, Comp.}, Monticello, IL, 2009, pp. 408-415.

\bibitem{MacKay99}
D. J. C. MacKay, 
``Good error correcting codes based on very sparse matrices,''
IEEE Trans. Inf. Theory, vol. 45, no. 2, pp. 399-431, Mar. 1999.




\bibitem{Sula17}
E. Sula,  J. Zhu,  A. Pastore,  S. H.Lim,  and  M. Gastpar, 
``Compute-forward multiple access (CFMA) with nested LDPC codes, ''
\textit{Proc. IEEE Int. Symp. Inf. Theory}, Aachen, Jun. 2017, pp. 2935-2939.




\bibitem{tiwh} S. Takabe, Y. Ishimatsu, T. Wadayama and M. Hayashi, ``Asymptotic analysis on spatial coupling coding for two-way relay channels,''
\textit{Proc. IEEE Int. Symp. Inf. Theory}, Vail, CO, June, 2018, pp. 1021-1025.

\bibitem{twh2} S. Takabe, T. Wadayama and M. Hayashi, ``Asymptotic analysis on spatial coupling coding for compute-and-forward relaying,'' arXiv:1807.01599, 2018.


\bibitem{Feng} {C. Feng, D. Silva and F. R. Kschischang, ``An algebraic approach to physical-layer network coding,'' 
IEEE Trans. Inf. Theory, vol. 59, no. 11, pp. 7576-7596, Nov. 2013.}

\bibitem{BICM} A. G. i F\`abregas, A. Martinez, and G. Caire, ``Bit-interleaved coded modulation,'' Found. Trends Commun. Inf. Theory, vol. 5, pp. 1-153, 2007.
\bibitem{Du} J. Du, L. Yang, J. Yuan, L. Zhou and X. He, 
``Bit mapping design for LDPC coded BICM schemes with binary physical-layer network coding,''
\textit{2016 IEEE Global Commun. Conf.}, Washington, DC, 2016, pp. 1-6.

\bibitem{Lei} L. Yang, T. Yang, J. Yuan, and J. An, 
``Achieving the near-capacity of two-way relay channels with modulation-coded physical-layer network coding,''
IEEE Trans. Wirel. Commun., vol. 14, no. 9, pp. 5225-5239, 2015.

\bibitem{Hou}
J. Hou, P. H. Siegel, L. B. Milstein and H. D. Pfister, 
``Capacity-approaching bandwidth-efficient coded modulation schemes based on low-density parity-check codes,'' 
IEEE Trans. Inf. Theory, vol. 49, no. 9, pp. 2141-2155, Sep. 2003.
\bibitem{Ash} A. Ashikhmin, G. Kramer and S. ten Brink, ``Extrinsic information transfer functions: model and erasure channel properties,'' IEEE Trans. Inf. Theory, vol. 50, no. 11, pp. 2657-2673, Nov. 2004.


\bibitem{Wang05}
C.-C. Wang, S.R. Kulkarni, and H.V. Poor,
``Density evolution for asymmetric memoryless channels, ''
IEEE Trans. Inf. Theory, vol. 51, no. 12, pp. 4216-4236, Dec. 2005.

\bibitem{Dana} Z. Faraji-Dana and P. Mitran, ``On non-binary constellations for channel-coded physical-layer network coding,''
 IEEE Trans. Wirel. Commun., vol. 12, no. 1, pp. 312-319, Jan. 2013.

\bibitem{Noori} M. Noori and M. Ardakani, ``On symbol mapping for binary physical-layer network coding with PSK modulation,'' 
IEEE Trans. Wirel. Commun., vol. 11, no. 1, pp. 22-26, Jan. 2012.

\bibitem{Wu} M. Wu, D. W\"ubben, and A. Dekorsy, ``Mutual information based analysis for physical-layer network coding with optimal phase control,''
\textit{9th Int. ITG Conference on Systems, Commun. Coding}, Munich, Germany, Jan. 2013, vol. 9, pp. 1-6.


\bibitem{Mezard}
M. M\'ezard and A. Montanari, 
\textit{Information, physics, and computation},
Oxford University press, 2009.
\bibitem{Yedla}
A. Yedla, P. S. Nguyen, H. D. Pfister, and K. R. Narayanan,
``Universal codes for the Gaussian MAC via spatial coupling,''
\textit{2011 49th Annual Allerton Conf. Commun., Control, Comp.}, Monticello, IL, 2011, pp. 1801-1808.


\end{thebibliography}
\end{document}